\documentclass[nohyperref]{article}

\usepackage{microtype}
\usepackage{graphicx}
\usepackage{subfigure}
\usepackage{booktabs} 

\usepackage{hyperref}

\usepackage[accepted]{icml2022}

\usepackage{amsmath}
\usepackage{amssymb}
\usepackage{mathtools}
\usepackage{amsthm}
\usepackage{enumitem}

\usepackage{amsmath,amsfonts,bm}
\newcommand\Tstrut{\rule{0pt}{2.3ex}}
\newcommand\Bstrut{\rule[-0.9ex]{0pt}{0pt}}

\newcommand{\FFN}{\mathrm{FFN}}

\newcommand{\set}[1]{\{#1\}}

\def\eqref#1{equation~\ref{#1}}

\def\1{\bm{1}}

\def\va{{\bm{a}}}
\def\vb{{\bm{b}}}
\def\vc{{\bm{c}}}

\def\vf{{\bm{f}}}
\def\vg{{\bm{g}}}
\def\vh{{\bm{h}}}

\def\vn{{\bm{n}}}

\def\vp{{\bm{p}}}
\def\vq{{\bm{q}}}

\def\vt{{\bm{t}}}
\def\vu{{\bm{u}}}

\def\vx{{\bm{x}}}

\def\vz{{\bm{z}}}

\def\mD{{\bm{D}}}

\def\mO{{\bm{O}}}

\def\mR{{\bm{R}}}
\def\mS{{\bm{S}}}

\def\mU{{\bm{U}}}

\def\mW{{\bm{W}}}
\def\mX{{\bm{X}}}

\DeclareMathAlphabet{\mathsfit}{\encodingdefault}{\sfdefault}{m}{sl}
\SetMathAlphabet{\mathsfit}{bold}{\encodingdefault}{\sfdefault}{bx}{n}

\def\gC{{\mathcal{C}}}

\def\gG{{\mathcal{G}}}

\def\gN{{\mathcal{N}}}

\newcommand{\softmax}{\mathrm{softmax}}

\allowdisplaybreaks

\usepackage[textsize=tiny]{todonotes}

\icmltitlerunning{Antibody-Antigen Docking and Design via Hierarchical Equivariant Refinement}

\begin{document}

\twocolumn[
\icmltitle{Antibody-antigen Docking and Design via Hierarchical Equivariant Refinement}



\icmlsetsymbol{equal}{*}

\begin{icmlauthorlist}
\icmlauthor{Wengong Jin}{broad}
\icmlauthor{Regina Barzilay}{mit}
\icmlauthor{Tommi Jaakkola}{mit}
\end{icmlauthorlist}

\icmlaffiliation{broad}{Eric and Wendy Schmidt Center, Broad Institute of MIT and Harvard}
\icmlaffiliation{mit}{CSAIL, Massachusetts Institute of Technology}

\icmlcorrespondingauthor{Wengong Jin}{wengong@csail.mit.edu}

\icmlkeywords{Generative Models, Computational Biology, Protein Design, Graph Neural Networks}

\vskip 0.3in
]



\printAffiliationsAndNotice{}  

\begin{abstract}
Computational antibody design seeks to automatically create an antibody that binds to an antigen. The binding affinity is governed by the 3D binding interface where antibody residues (paratope) closely interact with antigen residues (epitope). Thus, predicting 3D paratope-epitope complex (docking) is the key to finding the best paratope. In this paper, we propose a new model called Hierarchical Equivariant Refinement Network (HERN) for paratope docking and design. During docking, HERN employs a hierarchical message passing network to predict atomic forces and use them to refine a binding complex in an iterative, equivariant manner. During generation, its autoregressive decoder progressively docks generated paratopes and builds a geometric representation of the binding interface to guide the next residue choice. Our results show that HERN significantly outperforms prior state-of-the-art on paratope docking and design benchmarks.
\end{abstract}
\section{Introduction}

\begin{figure}
    \centering
    \includegraphics[width=0.48\textwidth]{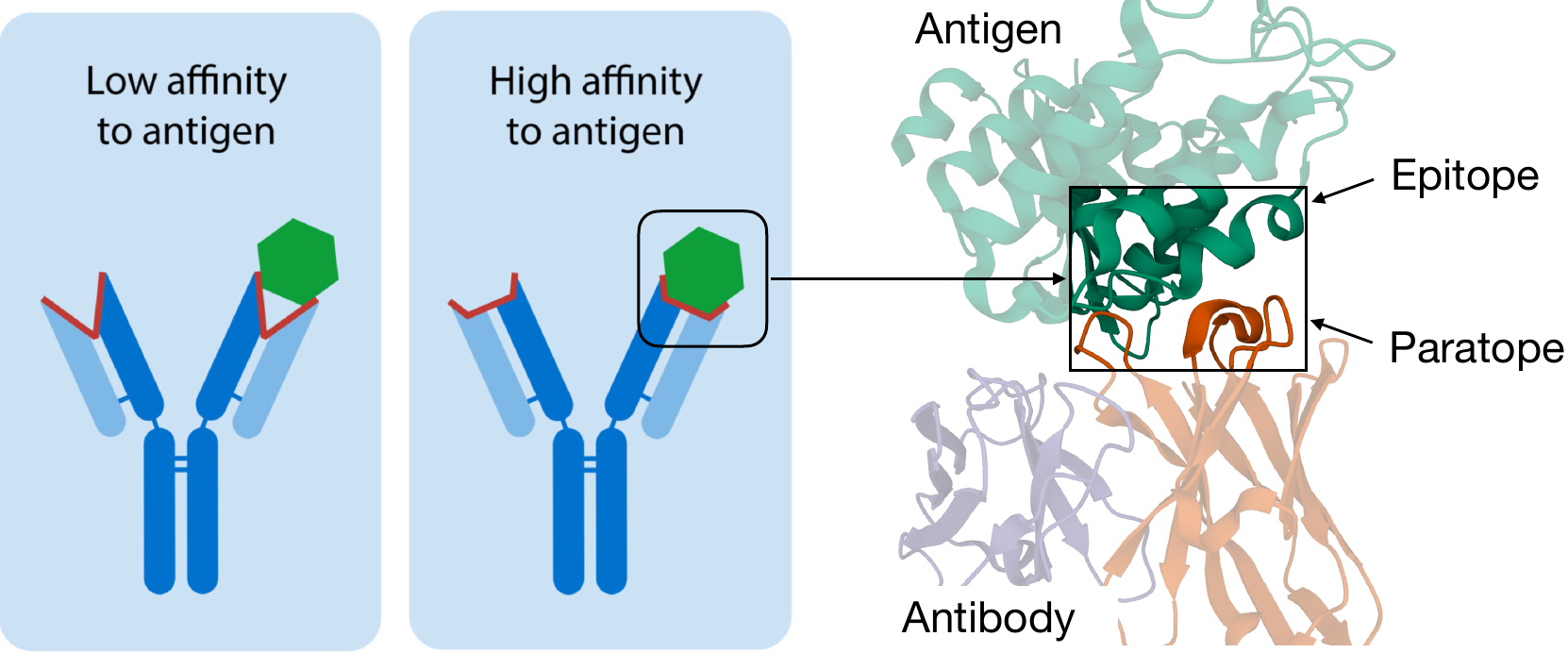}
    \vspace{-15pt}
    \caption{Illustration of paratope-epitope interaction. The binding affinity relies on the quality of paratope-epitope match. Figure partially adapted from online [\href{https://www.fluidic.com/resources/what-are-neutralizing-antibodies/}{link}].}
    \label{fig:shape}
    \vspace{-7pt}
\end{figure}

Recently, deep learning~\cite{shin2021protein,jin2021iterative} have demonstrated the feasibility of designing new antibodies to combat different pathogens (i.e., antigens). In this paper, we propose to further advance this capacity by tailoring antibody design models for binding a specific region of an antigen (an epitope). Figure~\ref{fig:shape} illustrates a binding interface consisting of an antigen epitope and an antibody paratope. Antibody binding affinity relies on the quality of paratope-epitope match, similar to a lock and a key. While a typical antigen has multiple epitopes, some of them may constitute more desirable targets from a therapeutic perspective. For instance, binding to a more conserved epitope decreases the chance of viral escape and prolongs antibody efficacy \cite{yuan2020highly}. Therefore, conditioning on a specific epitope gives us more control in antibody design.\footnote{In contrast to \citet{jin2021iterative}, we focus on binding interface design, a more general setting where bound antibody frameworks are not available. Our code is available at \url{github.com/wengong-jin/abdockgen}.}

\begin{figure}
    \centering
    \includegraphics[width=0.48\textwidth]{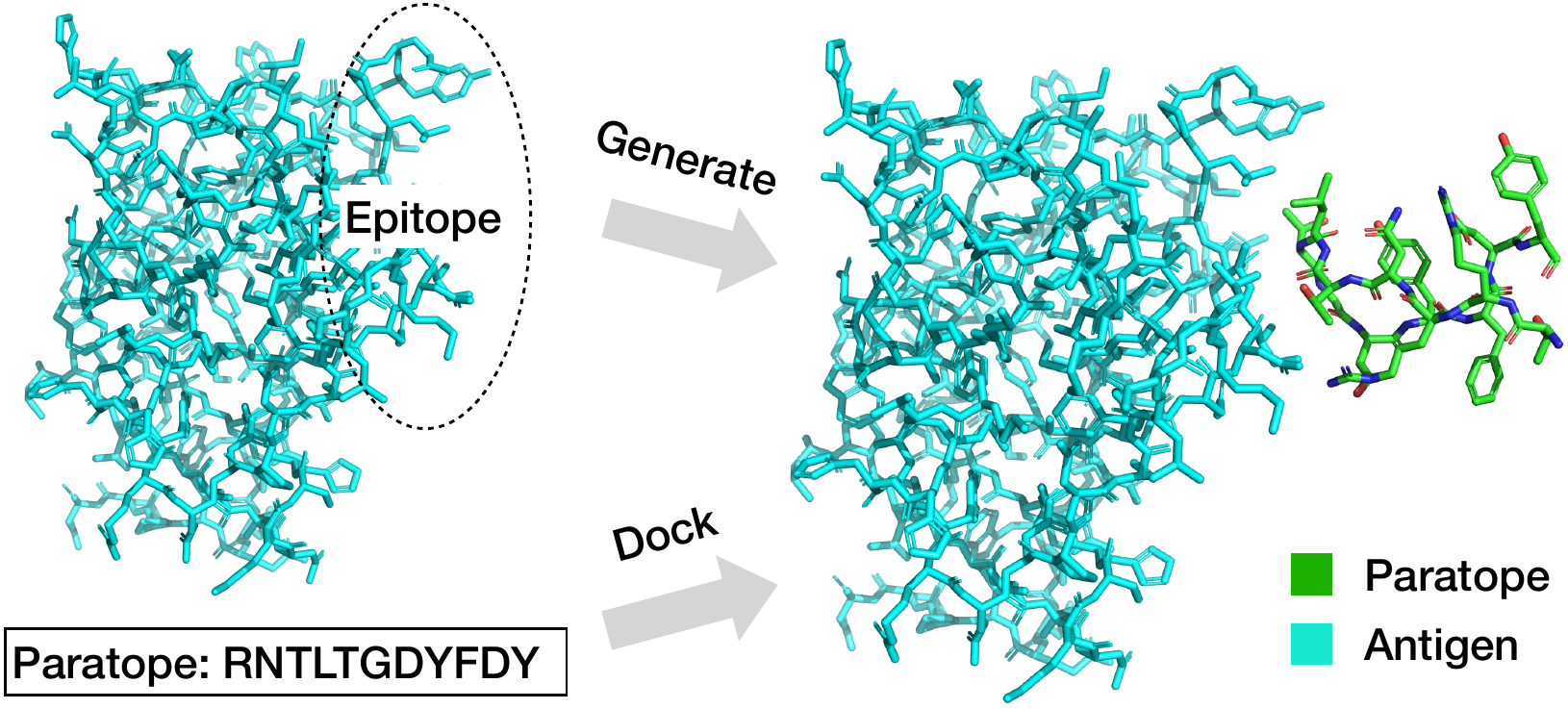}
    \vspace{-15pt}
    \caption{Illustration of paratope docking and design. Paratope design assumes the epitope is given by the user and seeks to generate a paratope sequence and its 3D structure for epitope binding. Paratope docking further assumes a paratope sequence is given and aims to predict the 3D coordinate of each atom.}
    \label{fig:formulation}
\end{figure}

From a computational perspective, epitope-specific antibody design can be decomposed into two complementary tasks: predicting the paratope-epitope complex for a paratope sequence (local docking)\footnote{Knowing the paratope-epitope 3D structure allows us to compute binding energy via physics-based models like Rosetta/FoldX.} and generating a paratope sequence that matches the shape of a given epitope. 
Both tasks can be formulated as point cloud completion (Figure~\ref{fig:formulation}), which imposes three modeling challenges.
The first is paratope flexibility. Existing rigid-body docking models~\cite{yan2020hdock,ganea2021independent} assume the ligand 3D structure is given and fixed, but such information is not available for a new designed paratope.
The second is equivariance. Previous work~\cite{jin2021iterative} cannot model paratope-epitope interaction because it predicts the absolute coordinates while the epitope can be rotated arbitrarily.
The third is efficiency. It is computationally expensive to build an all-atom structure from scratch because a binding interface usually involves hundreds of atoms. Previous architectures~\cite{ingraham2018learning} model protein backbone only and ignore the side-chain conformation, but side-chain atom contacts play an important role in binding.

In this paper, we propose a new architecture called Hierarchical Equivariant Refinement Network (HERN) to address the above challenges.
As opposed to rigid-body docking, HERN simultaneously folds and docks a paratope sequence and iteratively refines its 3D structure during generation. The main novelty of our approach is \emph{hierarchical equivariance}. HERN represents a binding interface as a dynamic, hierarchical graph. It consists of a residue-level graph with only C$_\alpha$ atoms and an atom-level graph that includes the side chains. This multi-scale representation allows us to factorize the paratope docking task into a global backbone update step and a local side-chain refinement step. To ensure equivariance of the docking procedure, HERN predicts the pseudo force between atoms instead of their absolute coordinates. During generation, HERN progressively docks paratope-epitope complex and uses its geometric representation to select the next amino acid.

We evaluate our method on standard paratope docking and design benchmarks~\cite{adolf2018rosettaantibodydesign}. In terms of docking, we compare HERN against HDOCK~\cite{yan2020hdock} and combined it with IgFold~\cite{ruffolo2022fast} for end-to-end paratope folding and docking. In terms of generation, we compare HERN against standard sequence-based generative models and a state-of-the-art structure-based methods~\cite{jin2021iterative}. HERN significantly outperformed all baselines in both settings, with 45\% absolute improvement in docking success rate.

\section{Related Work}

\textbf{Protein docking}.
Our method is closely related to protein-protein docking~\cite{kozakov2017cluspro,yan2020hdock,ganea2021independent}, which predicts the 3D structure of a protein-protein complex given the 3D structures of the two unbound proteins. These approaches assumes the paratope 3D structure is given and perform rigid-body docking. In contrast, HERN predicts the 3D structure of a paratope-epitope complex by simultaneously folding and docking a paratope sequence to an epitope. In that regard, our work is more similar to AF2-multimer~\cite{evans2021protein} that simultaneously folds two proteins. However, AF2-multimer relies on multi-sequence alignment (MSA) and sometimes template features to predict protein complex structures. Antibody paratopes are highly variable and generally lack MSA data. Moreover, existing docking models are not directly applicable to a generative setting where they need to dock a partial sequence for paratope design. Our approach does not require MSA data and can be easily integrated into a generative design workflow.

\textbf{Generative antibody/protein design.} 
Recently, the antibody design community started to explore deep generative models due to their computational efficiency compared to traditional physics-based models~\cite{lapidoth2015abdesign,adolf2018rosettaantibodydesign}. For example, \citet{liu2020antibody,shin2021protein,saka2021antibody,akbar2021silico} trained recurrent neural networks to generate paratope sequences. The limitation of sequence-based models is that they do not utilize the paratope 3D structure. Therefore, \citet{jin2021iterative} proposed a graph-based generative model that co-designs a paratope sequence and its 3D structure. However, none of these models incorporated the antigen structure. Thus, existing methods cannot design paratopes tailored for a specific epitope.

For general protein design, most previous methods are not conditioned on a target protein structure. While \citet{ingraham2019generative,strokach2020fast,karimi2020novo,cao2021fold2seq} proposed generative models conditioned on a backbone structure or protein fold, they do not model the interaction between a generated protein and its target protein. This limitation also applies to energy-based protein design models based on TrRosetta \cite{tischer2020design,norn2021protein}. While non-deep learning based methods \cite{adolf2018rosettaantibodydesign,cao2021robust} use docking algorithms to design target-specific proteins, they are computationally very expensive. In contrast, we propose a new generative model that leverages both 3D structural information and epitope structure.

\textbf{Protein structure encoder}. Prior work has utilized graph neural networks (GNN) \cite{fout2017protein}, 3D convolutional neural networks \cite{townshend2019end}, and equivariant neural networks (ENN) \cite{satorras2021n,eismann2021hierarchical} to encode protein 3D structures. Similar to our work, \citet{eismann2021hierarchical} developed a hierarchical ENN that represents a protein in terms of backbone (C$_\alpha$) and side chains. However, their method assumes the 3D structure of a full protein complex is given. Therefore, it can only be used to re-rank docked structures generated by another docking algorithm. In contrast, our model predicts the 3D structure of the backbone and side chains from scratch.
\citet{somnath2021multi} also developed a hierarchical GNN that represents a protein in terms of triangulated surfaces and residue C$_\alpha$ atoms. Their method is not applicable to our setting because it does not model the side chain atoms.

\section{Paratope Docking and Design with HERN}

When an antibody binds to an antigen, they form a joint structure called an antibody-antigen complex. Its binding interface consists of a \emph{paratope} and an \emph{epitope}. A paratope is a sequence of residues $\va = \va_1\cdots\va_n$ in the complementarity determining regions (CDRs) of an antibody. An epitope $\vb = \vb_1\vb_2\cdots\vb_m$ is composed of $m$ residues that are closest to a paratope. It is a subsequence of an antigen $\vc= \vc_1\vc_2\cdots\vc_M$, where $\vb_i = \vc_{e_i}$ and $e_i$ is the index of epitope residue $\vb_i$ in the antigen.

The epitope 3D structure $\gG_b$ is described as a point cloud of atoms $\set{\vb_{i,j}}_{1 \leq i \leq m, 1 \leq j \leq n_i}$, where $n_i$ is the number of atoms in residue $\vb_i$. Likewise, the 3D structure of a paratope and a paratope-epitope binding interface is denoted as $\gG_a$ and $\gG_{a,b}$, respectively. The first four atoms in any residue correspond to its backbone atoms (N, C$_\alpha$, C, O) and the rest are its side chain atoms. The 3D coordinate of an atom $\vb_{i,k}$ is denoted as $\vx(\vb_{i,k}) \in \mathbb{R}^3$. 

In this paper, we assume the 3D structure of the antigen and its epitope are given as inputs to our model. Hence, both paratope docking and generation can be formulated as a 3D point cloud completion task. Given a training set of antibody-antigen complexes, HERN learns to append an epitope structure $\gG_b$ with paratope atoms to form a binding interface $\gG_{a,b}$ (Figure~\ref{fig:formulation}).
Overall, HERN consists of three modules to model 3D paratope-epitope interaction for docking and generation.
\begin{itemize}[leftmargin=*,topsep=0pt,itemsep=0pt]
    \item Its \emph{encoder module} learns to encode a docked paratope-epitope complex into a set of atom and residue vectors.
    \item Its \emph{docking module} predicts the 3D structure of a paratope-epitope complex given any paratope sequence. In contrast to rigid body docking, our model simultaneously folds and docks a paratope sequence since the paratope 3D structure is unknown.
    \item Its \emph{decoder module} generates paratope amino acids autoregressively and runs the docking module to update the paratope-epitope complex in each generation step. This module is not used during paratope docking.
\end{itemize}

\subsection{Hierarchical Encoder}
\label{sec:hiermpn}

During docking and generation, the input to HERN is a point cloud with atoms from an antigen, its epitope, and its binding paratope. A point cloud typically contains over $10^3$ atoms so we need to encode it at different resolutions for computational efficiency. Specifically, we adopt a hierarchical message passing network (MPN) composed of an atom-level and a residue-level interface encoder. The atom-level interface encoder captures paratope-epitope interaction at the finest resolution, including atomic contacts in the side-chains. The residue-level encoder only considers backbone C$_\alpha$ atoms to focus on long-range residue interactions.

\subsubsection{Atom-level Interface Encoder}

The atom-level encoder keeps all the atoms in the point cloud $\gG_{a,b}$ and constructs a $K$ nearest neighbor graph $\gG_{a,b}^A$. Each node feature is a one-hot encoding of its atom name (e.g., N, C$_\alpha$, C$_\beta$, O). Each edge feature between two atoms $(\va_{i,k}, \vb_{j,l})$ is represented as
\begin{equation}
   \vf(\va_{i,k}, \vb_{j,l}) = \mathrm{RBF}(\Vert \vx(\va_{i,k}) - \vx(\vb_{j,l}) \Vert)
\end{equation}
where $\mathrm{RBF}(\cdot)$ encodes the distance between two atoms in a radial basis. We encode $\gG_{a,b}^A$ by a MPN which learns a vector representation $\vh(\va_{i,k}), \vh(\vb_{j,l})$ for each paratope atom $\va_{i,k}$ and epitope atom $\vb_{j,l}$.

\subsubsection{Residue-level Interface Encoder}

The residue-level encoder only keeps the C$_\alpha$ atoms and constructs a $K$ nearest neighbor graph $\gG_{a,b}^R$ at the residue level. Each residue is represented by a feature vector $\vf(\va_i)$ including its dihedral angles, polarity, hydropathy, volume, charge, and whether it is a hydrogen bond donor or acceptor. It then concatenates the amino acid features with the sum of atom-level encodings $\vh(\va_{i,k})$ within that residue as the initial residue representation
\begin{eqnarray}
    \tilde{\vf}(\va_i) &=& \vf(\va_i) \oplus \sum_k\nolimits \vh(\va_{i,k}) \\
    \tilde{\vf}(\vb_j) &=& \vf(\vb_j) \oplus \sum_l\nolimits \vh(\vb_{j,l})
\end{eqnarray}
Next, it builds a local coordinate frame for each residue and compute edge features $\vf(\va_i, \va_j)$ describing the distance, direction, and orientation between nearby residues~\citep{ingraham2019generative}. 
Lastly, the encoder takes the node and edge features into another MPN to compute the final representations $\set{\vh(\va_i)}, \set{\vh(\vb_j)}$ for paratope and epitope residues.

In summary, the output of our hierarchical encoder is a set of residue representations $\set{\vh(\va_i), \vh(\vb_j)}$ and atom representations $\set{\vh(\va_{i,k}), \vh(\vb_{j,l})}$. Crucially, the whole encoding process is rotation and translation invariant.

\begin{figure*}[t]
    \centering
    \includegraphics[width=\textwidth]{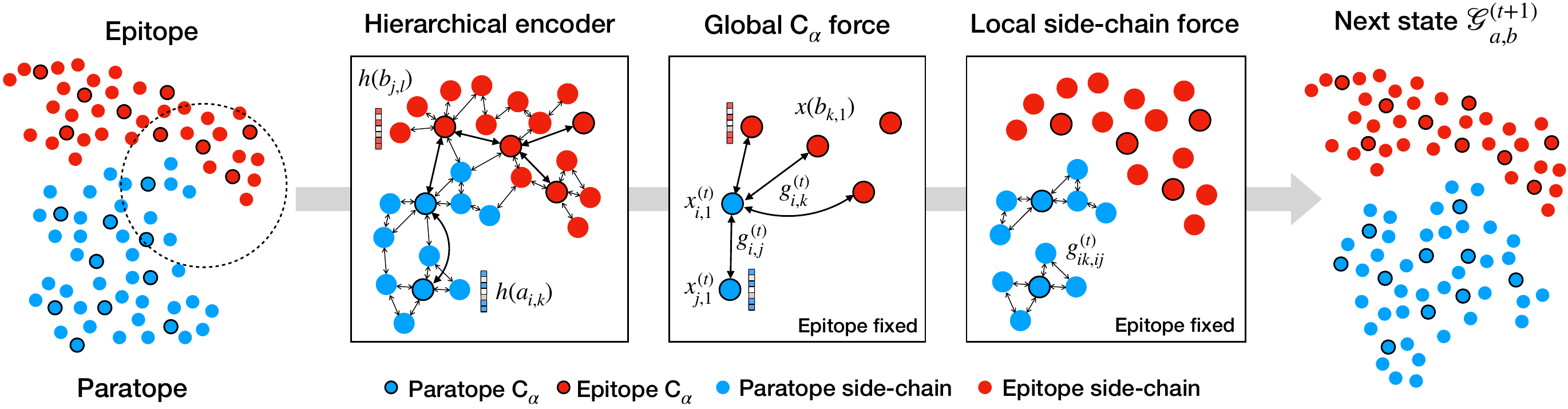}
    \vspace{-12pt}
    \caption{Illustration of the docking process. In each refinement step, the docking module first encodes the residues and atoms into vector representations. Next, it computes the residue-level force between C$_\alpha$ atoms and the local force between side-chain atoms within the same residue. Lastly, we update the paratope structure based on the predicted forces. Atom coordinates are projected to 2D for illustration.}
    \label{fig:docking}
\end{figure*}

\subsection{Docking Module}

Given a paratope sequence $\va$ and epitope 3D structure $\gG_b$, our docking module predicts the 3D coordinate of each atom to form a correct paratope-epitope complex $\gG_{a,b}$. The docking task is challenging as the model needs to simultaneously fold and dock the paratope. It is hard to predict the 3D structure in one shot due to the complex dependency between different variables $\vx(\va_{i,k}), \vx(\vb_{j,l})$ and various physical constraints. Thus, we propose an iterative refinement procedure to fold and dock a paratope onto a epitope.

\subsubsection{Initial atom coordinates}
\label{sec:initialize}

We investigate two strategies to initialize paratope coordinates at the outset of docking process.

\textbf{Random initialization}. We randomly initialize all coordinates by adding a small Gaussian noise around the center of the epitope ($\vb_{i,1}$ means the C$_\alpha$ atom of residue $\vb_i$)
\begin{equation}
    \vx^{(0)}(\va_{i,k}) = \frac{1}{m} \sum_i \vx(\vb_{i,1}) + \epsilon, \quad \epsilon \sim \mathcal{N}(0, 1),
\end{equation}
where the epitope center is defined as the mean of epitope C$_\alpha$ atoms. From now on, we will abbreviate paratope coordinates $\vx^{(0)}(\va_{i,k})$ as $\vx^{(0)}_{i,k}$ to simplify notation.

\textbf{Distance-based initialization}. Another strategy is to directly predict the pairwise distance $\mD \in \mathbb{R}^{(n+m)\times(n+m)}$ between paratope and epitope atoms and reconstruct atom coordinates from $\mD$.  
Specifically, let $\vh^{(0)}(\va_i)=\FFN(\vf(\va_{i}))$ be the initial representation of a paratope residue. We predict the pairwise distance as the following:
$$
\mD_{i,j} =
\begin{cases}
    \Vert \vh^{(0)}(\va_i) - \vh^{(0)}(\va_j) \Vert & i,j \leq n \\
    \Vert \vh^{(0)}(\va_i) - \vh(\vb_j) \Vert & i \leq n, j > n \\
    \Vert \vx(\vb_i) - \vx(\vb_j) \Vert & i, j > n\\
\end{cases}
$$
Intuitively, if $i$ belongs to the paratope ($i\leq n$) and $j$ to the epitope ($j > n$), $\mD_{i,j}$ is the Euclidean distance between $\vh_\mathrm{seq}(\va_i)$ and the antigen encoding $\vh(\vc_{e_j})$. Similarly, the distance between two paratope residues is modeled as the Euclidean distance between their sequence representations $\vh_\mathrm{seq}$. The distance between two epitope residues are directly calculated from their given coordinates $\vx(\vb_i)$.

Given this pairwise distance matrix $\mD$, we can obtain the 3D coordinates of each residue via eigenvalue decomposition of the following Gram matrix~\cite{crippen1978stable}
\begin{equation}
    \tilde{\mD}_{i,j} = 0.5(\mD_{i,1}^2 + \mD_{1,j}^2 - \mD_{i,j}^2), \quad \tilde{\mD} = \mU\mS\mU^\top
\end{equation}
where $\mS$ is a diagonal matrix with eigenvalues in descending order. The coordinate of each residue $\va_i$ is calculated as
\begin{equation}
    \tilde{\vx}^{(0)}_i = [\mX_{i,1}, \mX_{i,2}, \mX_{i,3}], \quad \mX = \mU\sqrt{\mS}.
\end{equation}
Note that the predicted coordinates $\set{\tilde{\vx}^{(0)}_i}$ retain the original distance $\mD$, but they are located in a different coordinate frame from the given epitope. Therefore, we apply the Kabsch algorithm \cite{kabsch1976solution} to find a rigid body transformation $(\mR, \vt)$ that aligns the predicted epitope coordinates $\set{\tilde{\vx}^{(0)}_{n+1}, \cdots, \tilde{\vx}^{(0)}_{n+m}}$ with the given epitope $\set{\vx(\vb_{n+1,1}), \cdots, \vx(\vb_{n+m,1})}$. Lastly, the C$_\alpha$ coordinates of each paratope residue is predicted as
\begin{equation}
    \vx^{(0)}(\va_{i,1}) = \vx^{(0)}_{i,1} = \mR \tilde{\vx}^{(0)}_{i} + \vt
\end{equation}
To predict the initial coordinates $\vx^{(0)}_{j,k}$ for other atoms $k$, we apply the same alignment procedure using epitope atoms $\set{\vx(\vb_{n+1,k}), \cdots, \vx(\vb_{n+m,k})}$ as the reference points.

\subsubsection{Hierarchical equivariant refinement}
\label{sec:refinement}

The initial coordinates are often inaccurate and needs to be refined into a correct structure. The refinement procedure must be equivariant because epitope coordinates are given and they can be rotated and translated arbitrarily. Inspired by force fields in physics, we propose to model the \emph{force} between atoms rather than predicting absolute coordinates \citep{jin2021iterative} to achieve equivariance.

Let $\gG_{a,b}^{(t)}$ be the predicted paratope-epitope complex in the $t$-th iteration, with predicted paratope coordinates $\set{\vx_{i,k}^{(t)}}$. In each iteration, we first encode $\gG_{a,b}^{(t)}$ using our hierarchical encoder to compute residue-level and atom-level representations $\vh^{(t)}(\cdot)$ for each residue. The force between two C$_\alpha$ atoms (atom index $k = 1$) is calculated based on their residue-level representation
\begin{eqnarray}
    \vg_{i,j}^{(t)} = \vg\big(\vh^{(t)}(\va_i), \vh^{(t)}(\va_j) \big) \cdot \big(\vx_{i,1}^{(t-1)} - \vx_{j,1}^{(t-1)}\big) \\
    \vg_{i,k}^{(t)} = \vg\big(\vh^{(t)}(\va_i), \vh^{(t)}(\vb_k)\big) \cdot \big(\vx_{i,1}^{(t-1)} - \vx(\vb_{k,1}) \big)
\end{eqnarray}
where $\vg$ is a feed-forward neural network with a scalar output. In practice, we truncate the magnitude of $\vg(\cdot)$ so that two atoms do not clash after the update.
We then update the C$_\alpha$ atom coordinates of each paratope residue by averaging the pairwise forces (Figure~\ref{fig:docking})
\begin{equation}
    \vx^{(t)}_{i,1} = \vx^{(t-1)}_{i,1} + \frac{1}{n}\sum_{j\neq i} \vg_{i,j}^{(t)} + \frac{1}{m} \sum_k \vg_{i,k}^{(t)} \label{eq:global-update}
\end{equation}
For other atoms types ($k\neq 1$), we only model the force between atoms in the same residue. It is computationally expensive to calculate forces between all atom pairs due to the large number of atoms in the binding interface. Specifically, the force from atom $\va_{i,j}$ to $\va_{i,k}$ is calculated from their atom-level representations
\begin{equation}
    \vg_{ik, ij}^{(t)} = \vg\big(\vh^{(t)}(\va_{i,j}), \vh^{(t)}(\va_{i,k}) \big) \big(\vx_{i,k}^{(t-1)} - \vx_{i,j}^{(t-1)}\big)
\end{equation}
Similarly, we truncate the force term if the atom distance is less than the Van der Waals radius. We update the atom coordinates by applying the local pairwise forces
\begin{equation}
    \vx_{i,k}^{(t)} = \vx_{i,k}^{(t-1)} + \frac{1}{n_i} \sum_{j} \vg_{ik, ij}^{(t)} \label{eq:local-update}
\end{equation}
In summary, the C$_\alpha$ update step in Eq.(\ref{eq:global-update}) focuses on the global structure refinement and the atom update step in Eq.(\ref{eq:local-update}) focuses on the local side-chain arrangement. Following the proof from \citet{satorras2021n}, it is easy to show that this coordinate update procedure is equivariant under epitope rotation and translation.

\subsection{Decoder Module}

\begin{figure}[t]
    \centering
    \includegraphics[width=0.48\textwidth]{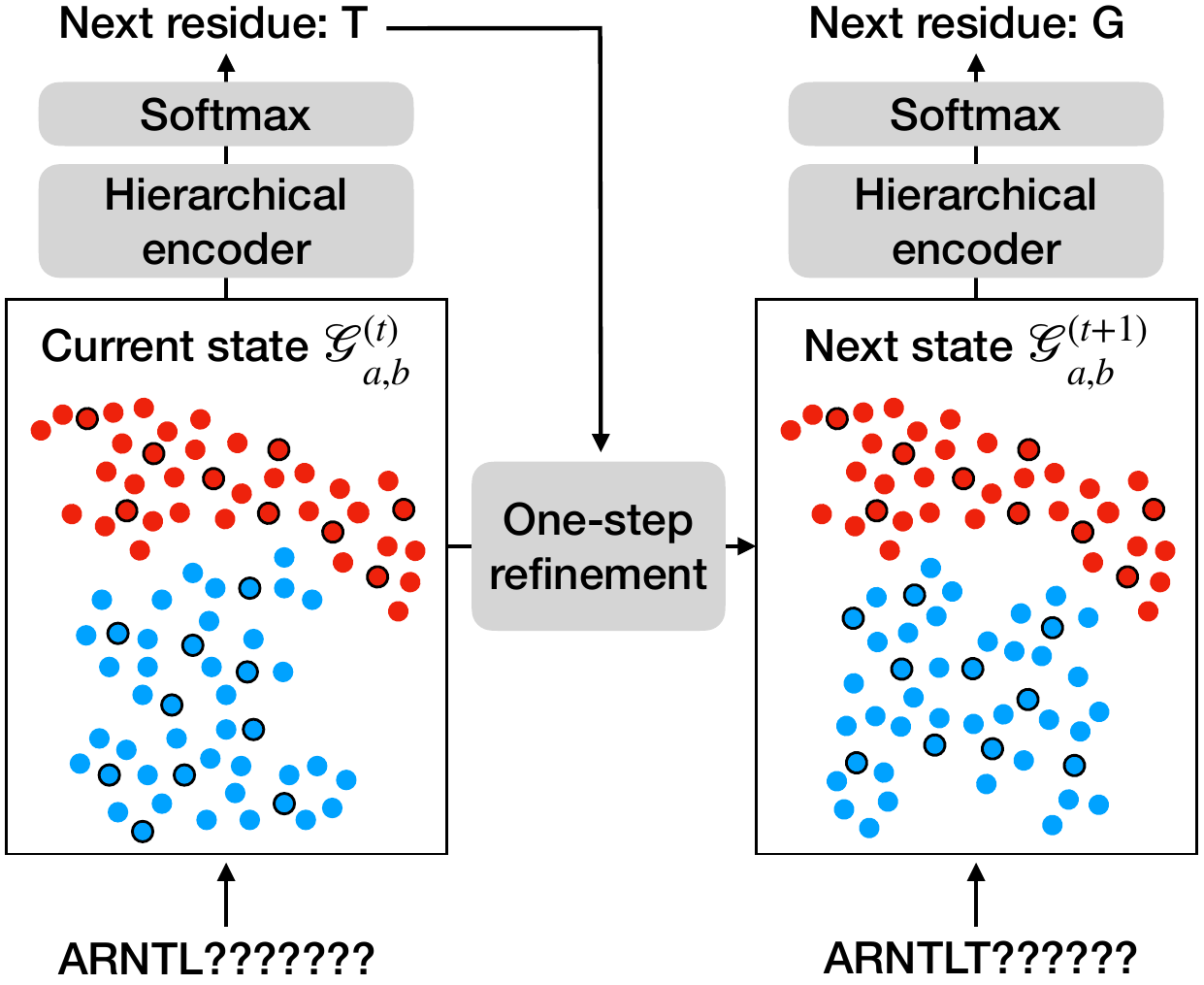}
    \vspace{-10pt}
    \caption{Illustration of the decoding process. In each generation step, HERN first encodes the predicted complex $\gG_{a,b}^{(t)}$ with a hierarchical encoder and predict the amino acid type of the next residue. It then use the docking module to perform one-step structure refinement to compute the next hidden state $\gG_{a,b}^{(t+1)}$.}
    \label{fig:decoder}
\end{figure}

Our decoder works together with the docking module to generate a paratope sequence autoregressively. The major difference between HERN and standard RNN decoders is the representation of intermediate states. In standard RNNs, an intermediate state is a partial paratope sequence. In contrast, the intermediate state in HERN is a paratope-epitope point cloud $\gG_{a,b}^{(t)}$ predicted by our docking module, which provides more structural information than a partial sequence. The overall generation process is illustrated in Figure~\ref{fig:decoder} and detailed below.

\subsubsection{Initial amino acid sequence}

To construct the initial state $\gG_{a,b}^{(0)}$, we need to first guess the amino acid type for every paratope residue because the docking module takes an entire sequence as input. Therefore, we predict the amino acid type of each paratope residue
\begin{align}
    \vp^{(0)}_{i} &= \softmax\big(\mW_0^\top \FFN(E_{\mathrm{pos}}(i)) \big)
\end{align}
where $\FFN(\cdot)$ is a feed-forward network and $E_{\mathrm{pos}}(i)$ is the positional encoding~\cite{vaswani2017attention} of paratope residue $\va_{i}$. $\vp^{(0)}_{i}[k]$ is the probability of residue $\va_{i}$ being amino acid type $k$. During generation, a paratope is represented as a sequence of  ``soft'' amino acids $(\vp^{(0)}_{1},\cdots,\vp^{(0)}_{n})$. This soft sequence is then combined with initial coordinates (section~\ref{sec:initialize}) to form the initial state $\gG_{a,b}^{(0)}$.

\subsubsection{State transition}

In each generation step $t$, HERN first encodes the predicted complex $\gG_{a,b}^{(t)}$ with the hierarchical encoder to learn a hidden representation for each residue $\vh^{(t)}(\va_i)$. Next, we predict the amino acid type of the next residue $\va_{t}$ by
\begin{equation}
    \vp_{t} = \softmax\big(\mW_s \vh^{(t)}(\va_{t})\big)
\end{equation}
During training, we apply teacher forcing and set $\va_{t}$ to its ground truth. During testing, we sample an amino acid type $k \sim \vp_{t}$ and set $\va_{t}=\mathrm{one\_hot}(k)$. The future residues remain as their initial soft assignment, i.e. the new paratope sequence becomes $(\va_1,\cdots,\va_t, \vp^{(0)}_{t+1},\cdots,\vp^{(0)}_{n})$. After a new residue is generated, we call the docking module to perform one structure refinement step to infer the new paratope-epitope structure $\gG_{a,b}^{(t+1)}$ (Figure~\ref{fig:decoder}). In other words, HERN co-designs a paratope sequence and its 3D structure by interleaving docking and generation steps.

Due to the soft assignment, there are two slight modifications in the docking module during paratope generation. First, the node feature in the hierarchical encoder is defined as an expectation $\vf(\va_{i}) = \sum_k \vp^{(0)}_{i}[k] \vf(k)$. Second, the docking module only predicts the four backbone atoms (N, C$_\alpha$, C, O) for the paratope since different amino acids have distinct side-chain atom configurations. Predicting all side-chain atoms would require a hard assignment of future residues during generation. Nonetheless, our hierarchical encoder still models all side-chain atoms in the epitope since their amino acid types are given.

\subsection{Training Loss}
\textbf{Docking loss}. 
The training loss of our docking module contains two terms. The first term comes from initial distance prediction, which is calculated as the Huber loss between the predicted distance $\mD_{i,j}$ and its ground truth.
The second term comes from the iterative refinement procedure. We compute the pairwise distance $\Vert \vx_{i,k}^{(t)} - \vx_{j,l}^{(t)} \Vert$ between all atoms in the binding interface and calculate the Huber loss compared to its ground truth in every refinement step. For training stability, we do not propagate the gradient across multiple refinement steps during back-propagation.

\textbf{Generation loss}. For paratope generation, our training loss includes another cross entropy term between the predicted probabilities $\vp_i^{(t)}$ and their ground truth amino acid types. During training, the docking module and decoder module are optimized simultaneously in one backward pass.
\section{Experiments}

\begin{table}[t]
\centering
\vspace{-5pt}
\caption{Paratope docking results with epitope size $m=20$.}
\vspace{5pt}
\begin{tabular}{lcc}
\hline
Method & DockQ & Success  \Tstrut\Bstrut \\
\hline
HDOCK & 0.237 & 48.3\% \Tstrut\Bstrut \\
\hline
HERN (random init) & \textbf{0.438} & \textbf{100\%} \Tstrut\Bstrut \\
HERN (distance init) & 0.377 & 87.9\%  \Tstrut\Bstrut \\
HERN (C$_\alpha$ only) & 0.357 & 89.7\% \Tstrut\Bstrut \\
HERN (no refinement) & 0.303 & 65.5\%   \Tstrut\Bstrut \\
\hline
\end{tabular}
\label{tab:docking}
\vspace{-5pt}
\end{table}

We evaluate HERN on paratope docking and generation tasks. For simplicity, we focus on the CDR-H3 paratope of a heavy chain, which is the major determinant of binding affinity. The corresponding epitope is constructed by the following procedure. For each antigen residue $\vc_i$, we first compute its shortest distance $d_i=\min_{j,k,l} \Vert \vx(\vc_{i,k}) - \vx(\va_{j,l}) \Vert$ to the paratope $\va$. We then rank the antigen residues based on $d_i$ and take the top $m$ residues as the epitope. The epitope sequence $\vb_1\cdots\vb_m=\vc_{e_1}\cdots \vc_{e_m}$ is sorted in the ascending order $e_1<\cdots<e_m$ so that it is a subsequence of an antigen and the order does not expose their proximity to the paratope.

\textbf{Data}. Our training data comes from the Structural Antibody Database (SAbDab)~\cite{dunbar2014sabdab}, which contains around 3K antibody-antigen complexes after filtering structures without antigens and removing duplicates. 
The test set for this task comes from \citet{adolf2018rosettaantibodydesign}. It contains 60 antibody-antigen complexes with diverse antigen types. For proper evaluation, we removed any paratope sequences from the training set if it is similar to one of the test set paratopes (sequence identity above 70\%). The training and validation set contains 2777 and 169 complexes, respectively. We use the same train/test split for docking and generation experiments.

\begin{table}[t]
\centering
\vspace{-5pt}
\caption{HERN docking performance with different epitope sizes.}
\vspace{5pt}
\begin{tabular}{lccc}
\hline
 & $m=20$ & $m=40$ & $m=80$ \Tstrut\Bstrut \\
\hline
DockQ & 0.438 & 0.413 & 0.375  \Tstrut\Bstrut \\
Success & 100\% & 96.6\% & 89.7\% \Tstrut\Bstrut \\
\hline
\end{tabular}
\label{tab:episize}
\vspace{-5pt}
\end{table}


\begin{figure*}[t]
    \centering
    \includegraphics[width=\textwidth]{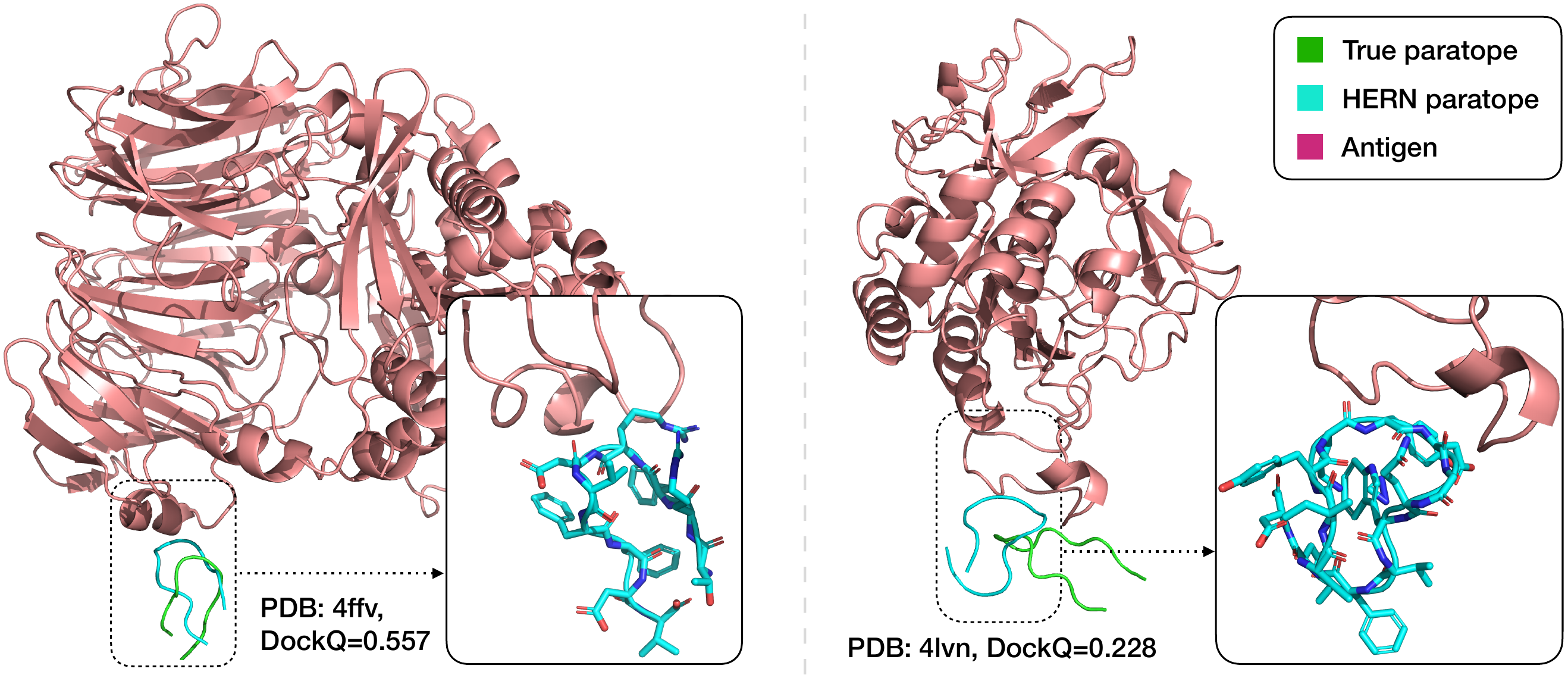}
    \vspace{-12pt}
    \caption{Comparison between the paratope structures predicted by HERN and the ground truth. The left figure presents a successful case with high DockQ score of 0.557. The right figure presents a failure case with low DockQ score of 0.228. The side chain structures of the docked paratopes are visualized in the right bottom corner. Best viewed in color.}
    \label{fig:structure}
\end{figure*}

\subsection{Paratope Docking}

In this task, the input to our model is an antigen 3D structure with a specified epitope $\vb_1\cdots\vb_m$ and a paratope sequence $\va_1\cdots\va_n$. The goal is to predict the 3D coordinates of all paratope atoms, namely the 3D paratope-epitope complex. This task is also called \emph{local docking}, as opposed to global docking which does not assume the epitope is given.

\textbf{Baselines}. We compare HERN with HDOCK~\cite{yan2020hdock}, a docking algorithm with state-of-the-art performance on protein docking benchmarks~\cite{ganea2021independent}. We choose HDOCK because it allows us to specify the receptor binding site and perform local docking.
Since HDOCK is a rigid-body docking software that requires a paratope unbound structure as input, we use IgFold~\cite{ruffolo2022fast} to predict the structure of an antibody heavy chain and CDR-H3. IgFold is a state-of-the-art antibody folding algorithm with comparable performance to AlphaFold~\cite{jumper2021highly}. Thus, we believe the combination of HDOCK and IgFold is a strong baseline for comparison.

\textbf{Metrics}. For all methods, we report the DockQ score~\cite{basu2016dockq}, a standard evaluation metric for docking. DockQ is a weighted average of three terms: contact accuracy, interface RMSD, and ligand RMSD. A docked structure is considered acceptable if its DockQ score is above 0.23. We define the success rate of a model as the percentage of acceptable docked structures.

\textbf{Hyperparameters}. 
Each MPN in our hierarchical encoder contains four message passing layers with a hidden dimension of 256. The docking module performs eight iterations of structure refinement and then calls OpenMM~\cite{eastman2017openmm} to refine the final structure. We use random initialization as the default scheme. All models are trained by an Adam optimizer for 20 epochs with 10\% dropout.

\textbf{Results}. We first compare all methods with epitope size $m=20$. As shown in Table~\ref{tab:docking}, HERN significantly outperforms HDOCK baseline in terms of average DockQ (0.237 vs 0.438) and success rate (48.3\% vs 100\%). In terms of speed, the average runtime for HERN is less than 1 second per instance, while HDOCK takes more than a minute. The reason is that HDOCK enumerates and ranks all possible docking poses (over 1000 candidates) in a brute-force manner. HERN instead treats docking as an optimization problem that requires only few iterations. 
In Figure~\ref{fig:structure}, we plot a successful case (PDB:4ffv) and a failure case (PDB:4lvn) docked by HERN. Indeed, the model receives a low DockQ score if a docked paratope has a different orientation. 

\subsubsection{Ablation Study}

Now we perform a series of ablation studies to investigate the importance of different model components.

\textbf{Epitope size}. First, we evaluate our method with increasing epitope sizes $m \in \set{20,40,80}$. Our results are summarized in Table~\ref{tab:episize}. Indeed, the average DockQ gradually decreases as the docking task becomes increasingly difficult, but the overall success rate remains high.

\textbf{Hierarchical encoder}. Second, we find that removing the atom-level interface encoder hurts the performance (C$_\alpha$ only, Table~\ref{tab:docking}), with DockQ decreasing from 0.438 to 0.357. These results demonstrate the necessity of utilizing side-chain information for docking.

\textbf{Structure refinement}. Third, we remove the iterative structure refinement procedure and run the docking module with distance-based initialization. Indeed, the model performance degrades significantly without structure refinement (DockQ: 0.438 vs 0.303, Table~\ref{tab:docking}), validating the benefit of equivariant structure refinement.

\textbf{Initialization schemes}. Lastly, we compare HERN performance under two coordinate initialization schemes. Interestingly, we find that random initialization performs better than distance-based initialization (0.438 v.s. 0.377, Table~\ref{tab:docking}). We hypothesize that random initialization serves as data augmentation and reduces over-fitting.

\subsection{Paratope Generation}

In this task, our model input contains only an antigen 3D structure with a specified epitope $\vb$. The goal is to generate a CDR-H3 paratope that binds to a given epitope. There are two notable differences between our setup and previous work on paratope design~\cite{jin2021iterative}. First, our model does not assume the framework region is given since our ultimate goal is \textit{de novo} antibody design. Second, previous methods are not conditioned on the antigen structure.

\textbf{Baselines}. In this task, we consider the following state-of-the-art baselines for comparison.
\begin{itemize}[leftmargin=*,topsep=0pt,itemsep=0pt]
\item \emph{RosettaAntibodyDesign} (RAbD)~\cite{adolf2018rosettaantibodydesign}: We apply their \textit{de novo} design protocol to design CDR-H3 paratopes. Starting from a random sequence, it performs 250 iterations of sequence design, docking, and energy minimization to find the best paratope with minimal interface energy. Different from other baselines, we provide RAbD with framework structure as input because it does not support CDR design without the framework.

\item \emph{Sequence model}: This model consists of an encoder and a decoder. The encoder is MPN whose input is a 3D antigen structure with a specified epitope. Its decoder is a RNN~\cite{lei2021attention} that generates a paratope sequence autoregressively. It also employs an attention layer to extract information from encoded epitope representations so that the generation process is conditioned on the epitope.

\item \emph{RefineGNN}~\cite{jin2021iterative} is a state-of-the-art generative model for antibody paratopes. Similar to HERN, its decoder operates on a 3D paratope structure. Since the original RefineGNN is conditioned on the framework region rather than the antigen, we replace its encoder by our antigen MPN encoder and use the attention layer to extract information from the epitope representation. We call this modified model \textit{CondRefineGNN} to make a distinction from its original architecture. 
\end{itemize}

\textbf{Metrics}. Following \citet{adolf2018rosettaantibodydesign}, we use amino acid recovery (AAR) as our metric. For a generated paratope $\tilde{\va}$, we define its AAR as $\sum_i \frac{1}{n} \mathbb{I}[\va_i = \tilde{\va}_i]$, the percentage of residues matching the corresponding residue in the ground truth $\va$. Specifically, for each epitope, we generate 100 paratope sequences and choose the one with the best log-likelihood to calculate AAR score. 

Given the critical role of contact residues in binding, we report another metric called contact AAR (CAAR). It is defined as $\sum_{i\in \gC} \frac{1}{\gC}\mathbb{I}[\va_i = \tilde{\va}_i]$, where $\gC$ is a set of residues making contact with the epitope. A residue $\va_i$ makes contact with the epitope if $\min_{j,k,l} \Vert \vx(\va_{i,k}) - \vx(\vb_{j,l}) \Vert < 4.0$.

\textbf{Hyperparameters}. 
Each MPN in our hierarchical encoder contains three message passing layers with a hidden dimension of 256. We run HERN under distance initialization by default. All models are trained by an Adam optimizer for 10 epochs with 10\% dropout.

\textbf{Results}. As shown in Table~\ref{tab:design}, HERN achieves the best AAR and CAAR score on this benchmark. Indeed, Cond-RefineGNN yields sub-optimal performance because it does not model the 3D paratope-epitope complex. The paratope 3D structure alone is not sufficient for reconstructing the true paratope. The sequence model yields even lower AAR because it does not model the paratope 3D structure at all.

\subsubsection{Ablation study}

We further perform two ablation studies to study the importance of initialization and hierarchical encoder in the context of paratope generation.

\textbf{Hierarchical encoder}. When removing the atom-level interface encoder (C$_\alpha$ only), the generation performance degrades substantially (Table~\ref{tab:design}), with AAR decreasing from 34.1\% to 30.4\% under distance-based initialization. Indeed, both docking and generation experiments support the importance of hierarchical encoding.

\textbf{Initialization}. Unlike docking, HERN generation performance becomes much worse under the random initialization (34.1\% vs 31.0\%, Table~\ref{tab:design}). This result is expected because the paratope structure looks almost random in the beginning and does not provide any useful information for predicting the first few amino acids. This suggests that we need different initialization schemes for docking and generation.

\begin{table}[t]
\centering
\vspace{-5pt}
\caption{CDR-H3 paratope generation results with $m=20$. }
\vspace{5pt}
\begin{tabular}{lcccc}
\hline
Method & AAR & CAAR \Tstrut\Bstrut \\
\hline
RAbD & 28.6\% & 14.9\% \Tstrut\Bstrut \\
Sequence decoder & 32.2\% & 17.6\% \Tstrut\Bstrut \\
CondRefineGNN & 33.2\% & 19.7\% \Tstrut\Bstrut \\
\hline
HERN (distance init) & \textbf{34.1\%} & \textbf{20.8\%} \Tstrut\Bstrut \\
HERN (random init) & 31.0\% & 15.6\% \Tstrut\Bstrut \\
HERN (C$_\alpha$ only) & 30.4\% & 19.0\% \Tstrut\Bstrut \\
\hline
\end{tabular}
\label{tab:design}
\end{table}

\subsubsection{Interaction energy}

The ultimate goal of antibody design is to discover better paratopes rather than reconstructing known binders. Since HERN generates a full paratope-epitope complex, we can compare a designed paratope with its native one in terms of interaction energy. For a native paratope, we use OpenMM to refine its crystal structure and apply FoldX \cite{delgado2019foldx} to calculate its interaction energy $E_\mathrm{true}$. For a HERN-designed paratope, we calculate its energy $E_\mathrm{design}$ in a similar way but using its docked structure. The improvement of a designed paratope is defined as $\Delta E = E_\mathrm{design} - E_\mathrm{true}$, which is the lower the better.\footnote{We note that $\Delta E$ is an approximation of true binding energy because $E_\mathrm{design}$ is calculated over a docked structure.}

\begin{figure*}[t]
    \centering
    \includegraphics[width=\textwidth]{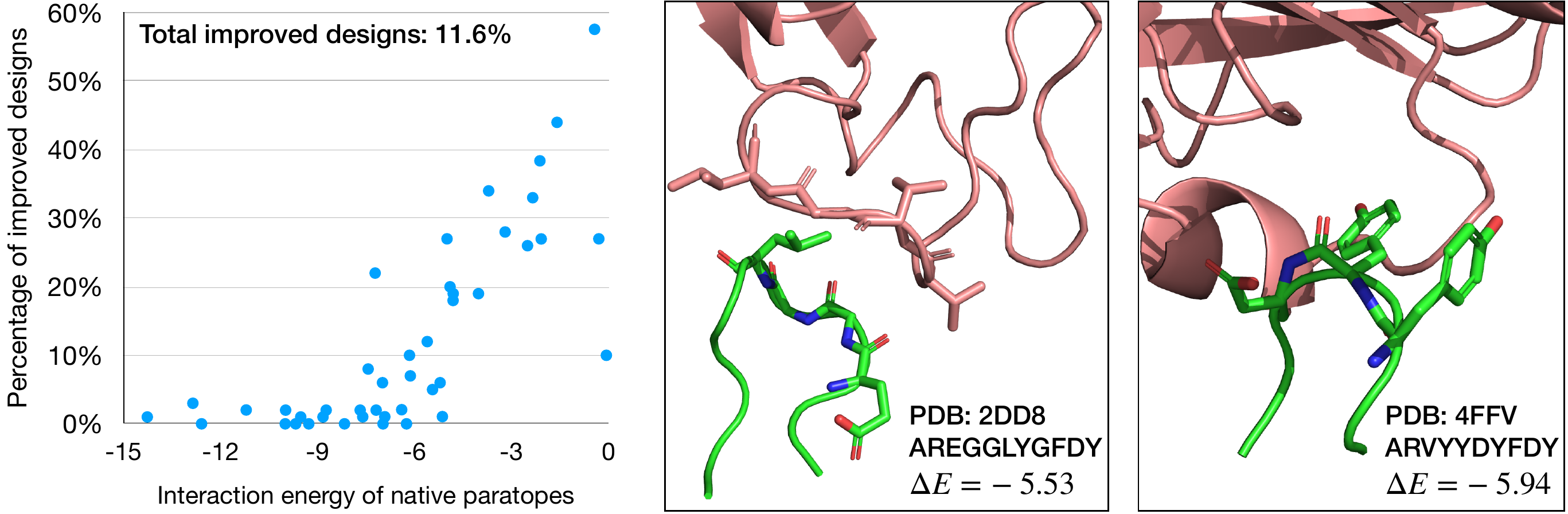}
    \vspace{-12pt}
    \caption{Left: Percentage of designed paratopes with improved interaction energy. The x-axis represent the interaction energy of a native paratope and the y-axis is the percentage of paratopes with lower interaction energy than the native ones (among 100 generated candidates). Right: Two examples of designed paratope sequences with lower interaction energy.}
    \label{fig:energy}
\end{figure*}

\begin{figure*}[t]
    \centering
    \includegraphics[width=\textwidth]{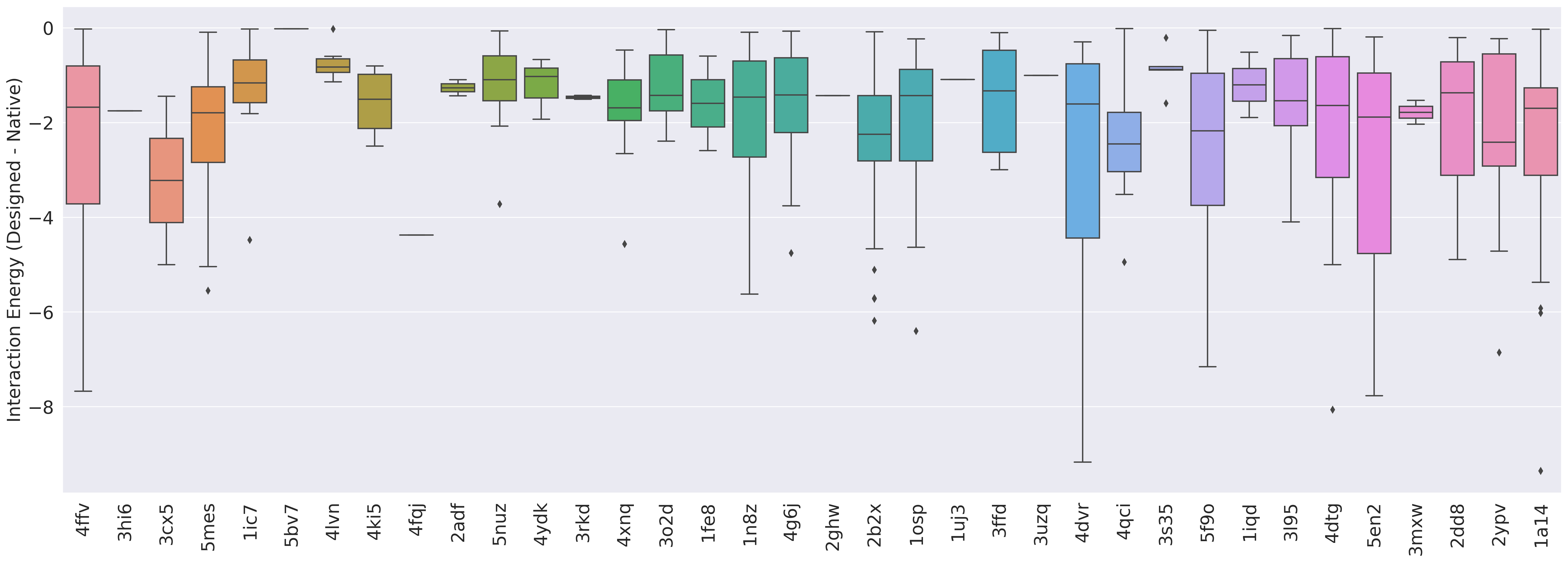}
    \vspace{-15pt}
    \caption{Distribution of $\Delta E$ among the 11.6\% improved paratopes. The y-axis represents the improvement $\Delta E = E_\mathrm{design} - E_\mathrm{true}$, which is the lower the better. A test case is not shown if none of the designed paratopes have lower energy.}
    \label{fig:improvement}
\end{figure*}

To test whether HERN is able to discover better paratopes, we generate 100 paratope sequences for each test epitope and calculate their $\Delta E$. As shown in Figure~\ref{fig:energy}, 11.6\% of generated paratopes achieved lower (better) interaction energy than the native binders. In general, HERN produces more improved designs when $E_\mathrm{true}$ is higher, but it is able to generate better paratopes in some hard cases with native interaction energy $E_\mathrm{true} < -10$. Figure~\ref{fig:energy} shows two examples of designed paratopes (docked structure) with high improvement. Figure~\ref{fig:improvement} further shows the distribution of $\Delta E$ among 11.6\% improved paratopes for each test case. Overall, HERN is able to discover better paratopes across many epitopes but there are plenty room for improvement. 

\section{Conclusion}
In this paper, we have developed a hierarchical, equivariant architecture for paratope docking and design. Our model significantly outperforms existing docking methods with a wide margin while running orders of magnitudes faster. Moreover, we demonstrate that HERN is able to design new paratopes with better binding energies than native paratopes. 
Indeed, there remain various ways to improve our model. First, HERN assumes the epitope is given as input. It needs to be combined with epitope prediction approaches~\cite{del2021neural} for end-to-end antibody engineering. Second, our current evaluation focus on the design of CDR-H3 paratopes as a proof of concept. We will extend our experiments to design all six CDRs jointly. Lastly, we plan to combine HERN with reinforcement learning to find paratopes with optimal interaction energy.

\section*{Acknowledgement}
We would like to thank Rachel Wu and Jeremy Wohlwend for their valuable feedback on the manuscript. This work is supported by a grant from Sanofi on next generation molecular design, DTRA Discovery of Medical Countermeasures Against New and Emerging (DOMANE) threats program, DARPA Accelerated Molecular Discovery program, Eric and Wendy Schmidt Center at the Broad Institute, and Abdul Latif Jameel Clinic for Machine Learning in Health.

\nocite{langley00}

\bibliography{main}
\bibliographystyle{icml2022}

\newpage
\appendix
\onecolumn
\section{Model Architecture Details}

\textbf{Amino acid features.} Each amino acid is represented by six features: polarity $f_p\in\set{0,1}$, hydropathy $f_h\in [-4.5, 4.5]$, volume $f_v\in [60.1, 227.8]$, charge $f_c\in \set{-1,0,1}$, and whether it is a hydrogen bond donor $f_d\in\set{0,1}$ or acceptor $f_a\in\set{0,1}$. For hydropathy and volume features, we expand it into radial basis with interval size 0.1 and 10, respectively. As a result, the amino acid feature dimension is 112.

\textbf{Residue-level edge features.}
For each residue $\va_i$, we define its local coordinate frame $\mO_i = [\vc_i, \vn_i, \vc_i \times \vn_i]$ as
\begin{equation}
    \vu_i = \frac{\vx_i - \vx_{i-1}}{\lVert \vx_i - \vx_{i-1} \rVert}, \quad
    \vc_i = \frac{\vu_i - \vu_{i+1}}{\lVert \vu_i - \vu_{i+1} \rVert}, \quad
    \vn_i = \frac{\vu_i \times \vu_{i+1}}{\lVert \vu_i \times \vu_{i+1} \rVert}
\end{equation}
Based on the local frame, we compute the following edge features
\begin{equation}
    \vf(\va_i, \va_j) = \bigg(
    E_{\mathrm{pos}}(i - j), \quad
    \mathrm{RBF}(\Vert \vx_{i,1} - \vx_{j,1} \Vert),  \quad
    \mO_i^\top \frac{\vx_{j,1} - \vx_{i,1}}{\Vert \vx_{i,1} - \vx_{j,1} \Vert},\quad
    \vq(\mO_i^\top \mO_j)
    \bigg). \label{eq:edge-feature}
\end{equation}
The edge feature $\vf_{ij}$ contains four parts. The positional encoding $E_{\mathrm{pos}}(i - j)$ encodes the relative distance between two residues in an antibody sequence. 
The second term $\mathrm{RBF}(\cdot)$ is a \emph{distance} encoding lifted into radial basis. 
The third term in $\vf_{ij}$ is a \emph{direction} encoding that corresponds to the relative direction of $\vx_j$ in the local frame of residue $i$. 
The last term $\vq(\mO_i^\top \mO_j)$ is the \emph{orientation} encoding of the quaternion representation $\vq(\cdot)$ of the spatial rotation matrix $\mO_i^\top \mO_j$.

\textbf{MPN architecture.} Our MPN contains $L$ message passing layers. Let $\mathcal{N}_i$ be the set of neighbor nodes for residue $\va_i$. Each MPN layer consists of a standard message passing step followed by an aggregation step with residual connection, where $\vh_0(\va_i) = \vz(\va_i)$.
\begin{equation}
    \vh_{l+1}(\va_i) = \vh_l(\va_i) + \sum_{j\in \gN_i} \FFN \big(\vh_l(\va_i), \vh_l(\va_j), \vf(\va_j), \vf(\va_i, \va_j)\big) \;\; (0 \leq l < L)
\end{equation}

\end{document}